\documentclass[12pt]{article}

\usepackage{setspace} % spacing

\usepackage{amsmath}
\usepackage{amssymb}
\usepackage{breqn}
\usepackage{authblk}
\usepackage{subcaption}
\usepackage{booktabs}
\usepackage{longtable,lscape,rotating}
\usepackage{adjustbox}
\usepackage{fullpage}
\usepackage{graphicx}
\usepackage{float}
\usepackage{multirow}
\usepackage{booktabs}
\usepackage{colortbl, xcolor}
\usepackage[bottom]{footmisc}
\usepackage{tikz}
\usetikzlibrary{positioning}
\usepackage{hyperref}

\usepackage{comment}
\usepackage{listings}

 \usepackage[natbib=true,backend=biber,sorting=nyt,style=apa]{biblatex}

\newcommand\E{\mathbb{E}}

\newcommand\indep{\perp\!\!\!\perp}

\title{Post-treatment problems: What can we say about the effect of a treatment among sub-groups who (would) respond in some way?}

\author[1,2]{Chad Hazlett}
\author[3]{Nina McMurry}
\author[1]{Tanvi Shinkre}

\affil[1]{Department of Political Science, UCLA}
\affil[2]{Department of Statistics and Data Science, UCLA}
\affil[3]{Department of Political Science, Vanderbilt University}
\begin{document}
%TC:ignore

\maketitle

\begin{abstract}
Investigators are often interested in how a treatment affects an outcome for units responding to treatment in a certain way. We may wish to know the effect among units that, for example, meaningfully implemented an intervention, passed an attention check, or demonstrated some important mechanistic response. Simply conditioning on the observed value of the post-treatment variable introduces problematic biases. Further, the identification assumptions required of several existing strategies are often indefensible. We propose the Treatment Reactive Average Causal Effect (TRACE), which we define as the total effect of treatment in the group that, if treated, would realize a particular value of the relevant post-treatment variable. By reasoning about the effect among the “non-reactive” group, we can identify and estimate the range of plausible values for the TRACE. We demonstrate the use of this approach with three examples: (i) learning the effect of police-perceived race on police violence during traffic stops, a case where point identification may be possible; (ii) estimating effects of a community policing intervention in Liberia, in communities that  meaningfully implemented it, and (iii) studying how in-person canvassing affects support for transgender rights, among participants for whom the intervention would result in more positive feelings towards transgender people.
\end{abstract}
%TC:endignore

\section{Introduction}
The effect of a treatment within groups defined by a response to treatment is often of  great interest. For example, we may be interested in the average effect among those who complied with a treatment, received a well-implemented treatment, were attentive, or demonstrated some other mechanistic reaction considered important or even a pre-requisite for a treatment effect. These ``effect among those who..." questions may be of primary research interest, or may arise when no effect was detected on average but where we recognize that implementation challenges, low compliance, or inattentive participants may have attenuated our estimate. 

While many investigators have attempted to examine effects in sub-groups defined by post-treatment variables, it is well understood that even in randomized trials, restricting the sample in this way compromises treatment effect estimates, including for that sub-group \citep{montgomery2018conditioning}.\footnote{The term ``post-treatment'' denotes here any variable that may be influenced by treatment. Variables measured after treatment assignment are not necessarily post-treatment, if they cannot be affected by treatment.}

We propose a straightforward approach for making bounded inferences about the “Treatment-Reactive Average Causal Effect” (TRACE), the total effect within the sub-group that, if treated, would have realized a specified value of a post-treatment variable. This differs from existing estimands such as the survivor average causal effect (SACE), mediation-related quantities (e.g. the controlled direct effect, CDE), “as-treated” or “per-protocol” estimands, and the local average treatment effect (LATE). The resemblance between the TRACE and the LATE is of particular interest to those familiar with the instrumental variables (IV) framework. As we discuss below, the most important distinction is that while IV requires the exclusion restriction \textemdash meaning here that treatment can have no ``direct effect'' on the outcome \textemdash we are interested in a total effect of treatment on the outcome, which includes that direct effect.

Our identification analysis begins by describing sharp “trimming” bounds on the TRACE, which require no additional assumptions. Potentially tighter trimming bounds are available when (i) the relevant post-treatment variable is observed for all units and (ii) the investigator is willing to assume monotonicity of this variable with respect to treatment. In many applications, however, these bounds remain wide. Our primary analysis therefore considers the potential 
to reach more informative conclusions by reasoning about another quantity:  the treatment effect among units that would \emph{not} respond to treatment, TRACE(0). For example, investigators may argue about the sign of TRACE(0), that it is near zero, or that its magnitude cannot exceed the TRACE itself. Such restrictions algebraically limit the possible values of TRACE. These results are combined with the trimming bounds to ensure no tighter range can be obtained without introducing additional assumptions. As we illustrate, assumptions on the range of TRACE(0) that may be defensible in the study context can lead to substantially informative conclusions. Additionally, in circumstances where a mediating event must occur in order for the outcome event to occur, it is sometimes plausible to argue that $\text{TRACE}(0)=0$, leading to point identification of the TRACE.  

We illustrate our approach, first, by demonstrating how a small number of sample statistics could allow point identification of the effect of police-perceived race on police violence during traffic stops, an area where post-treatment selection into the data has been a serious problem \citep{knox_administrative_2020}. We then analyze a randomized trial of a community policing program that produced largely null average effects, but in which investigators observed limited implementation of key components of the prescribed intervention \citep{morse_strengthening_2024}. This illustrates a valuable use-case, helping to reason about whether a null overall effect reflects low implementation as opposed to a small effect even where well implemented. Finally, we examine the persuasive effects of a 10-minute conversation on support for transgender rights, among participants whose feelings toward transgender individuals became more favorable following the intervention \citep{broockman_durably_2016}.

\section{Background}

\subsection{The problem}

\begin{figure}[htp!]
\begin{center}
\begin{tikzpicture}[->, thick]
\node (d) at (0,0) {$D$};
\node (y) at (4,0) {$Y$};
\node (m) at (2,0) {$M$};
\node (u) at (3.75,1) {$U$};
\node (x) at (2.25,1) {$X$};
\path (d) edge (m);
\path (m) edge (y);
\draw (d) to [bend right=30] (y);
\path[dashed] (u) edge (m);
\path[dashed] (u) edge (y);
\path (x) edge (m);
\path (x) edge (y);
\end{tikzpicture}
\caption{\textit{Causal structure of concern.} A treatment ($D$), which is unconfounded with the outcome of interest ($Y$) may affect $M$ in some sub-group. While some confounders of $M$ and $Y$ may be measured ($X$), we cannot rule out the presence of unobserved common causes of $M$ and $Y$ ($U$).}\label{fig:trace-dag}
\end{center}
\end{figure}
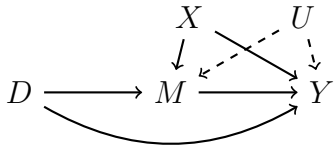

Consider the graph in Figure \ref{fig:trace-dag}. For concreteness, and previewing one of our applications, let $D$ be the treatment assignment for a community policing program, $M$ be some measure of program implementation, and $Y$ be an outcome of interest, such as crime rates. If we were interested in the effect of the program on the outcome, it would be natural to compare the outcomes in the treated group ($D=1$) with the outcomes in the control group ($D = 0$). However, the treated group potentially includes communities that did not actually implement the program, likely attenuating the treatment effect estimate.

A natural temptation is to look at only the group that implemented the program effectively, i.e. conditioning on $M$. This is problematic; first, since $M$ is post-treatment, conditioning on it blocks part of the effect of $D$ on $Y$. Second, it introduces a bias. We cannot typically rule out unobserved confounders of $M$ and $Y$ (shown as $U$). Because $M$ is a consequence of both $D$ and $U$, it is a collider \citep{pearl2009causality}, and conditioning on it can create an association between $D$ and $U$, leading $D$ and $Y$ to become associated (via $D \leftarrow U \rightarrow Y$) for reasons other than the causal effect of $D$. 

To avoid confusion when comparing this approach to IV below, note that the estimand
of interest here is the effect of the treatment ($D$) on the outcome ($Y$) within a sub-group 
related to $M$. While the causal structure resembles that assumed in the IV framework, in that setting the effect of interest is that of $M$ on $Y$, while $D$ plays the role of instrument. That is, the total effect in our setting corresponds to the intent-to-treat or reduced-form effect in the analogous IV setting. 

\subsection{Examples}

This arrangement appears in a wide variety of settings we organize into four broad types, all sharing the causal structure shown in Figure~\ref{fig:trace-dag}.

\paragraph{Type 1. Manipulation, attention, and implementation checks.}In these cases, $M$ represents a variable that, while not necessary for an effect of $D$ on $Y$, is thought to be of central importance for this effect. This encompasses our motivating example, where realization of the program on the ground ($M$) is thought to be important for its effect on $Y$. Other examples include cases where $M$ measures participants' compliance with a protocol, or responses to manipulation checks or attention screens performed after the treatment.\footnote{Note that in some cases, $M$ may only be measurable in the treated group. For example, if the program entails facilitated community meetings with police, a variable like ``attendance'' at such meetings may be poorly defined in the control group. Such $M$ variables are difficult to integrate into conventional analyses but can still be utilized for the TRACE.} 

\paragraph{Type 2. Mechanism of interest.}Here $M$ may not be the primary or only way for $D$ to affect $Y$, but it is of theoretical interest as part of a mechanism under investigation. For example, in our analysis of \citet{broockman_durably_2016} below, we wish to examine the effect of a perspective-taking intervention on support for transgender non-discrimination policies among participants who react to the intervention with increased warmth toward transgender individuals. This setting resembles questions in the mediation literature regarding direct vs. indirect effects, but as we describe in Section~\ref{subsec:compareestimands}, the TRACE engages a different causal question. Our approach also avoids problematic assumptions regarding the absence of mediator-outcome confounders, albeit at the cost of partial rather than point identification.

\paragraph{Type 3. ``Necessary condition'' mediators.}In this setting, $M$ is a necessary condition for $Y$ to possibly occur (equal 1). An example can be found in studies of the effect of a driver's race ($D$), as perceived by police, on the potential for police violence ($Y$) during a traffic stop ($M$). The chance of a traffic stop occurring ($M=1$) may be itself a function of police-perceived race. $Y$ can be defined specifically as ``police violence during a traffic stop,'' ensuring that no police violence of this kind can occur without a stop, i.e. $Y=0$ if $M=0$. Although this is merely an extreme case of $M$ being a mediator, we give it a separate category because the necessity of $M=1$ for $Pr(Y=1)>0$ can lead to point identification, as detailed below. 

\paragraph{Type 4. Treatment-responsive measurement of $M$.}In another setting, we may wish to know the effect of $D$ on $Y$ among those with some value of $M$, where measurement of $M$ is affected by $D$.  For example, suppose we wish to know the effect of cancer screening on survival, among those who would have received a positive result for cancer in \textit{that} screening (ruling out always-takers). The result here depends on our ability to reason about the effect of screening on survival for patients who, if screened, would have screened negative, on its own or in comparison to that in the group who would screen positive. This engages a complicated set of concerns about the effects of screening on behaviors, the false negative rate, potential harms due to false positives, and other challenges that we examine elsewhere.   

\section{Proposal}

\subsection{Setup and Notation}
\label{notation}
Continuing with the notation defined above, let $Y_i(d, m)$ denote the potential value of $Y$ for unit $i$ with treatment status $D_i=d$ and mediator value $M_i=m$. Let $M_i(d)$ denote the ``potential mediator,'' the value of $M$ for unit $i$ under treatment status $d$ \citep{imai_identification_2010, imai_comment_2014}. By writing potential outcomes, we assume consistency for both the mediator and the outcome: when $D_i=d$ and $M_i=m$ are observed, the observed outcome satisfies $Y_i = Y_i(d, m)$ and the observed mediator value satisfies $M_i = M_i(d)$.

In the following analysis, we need only consider two potential outcomes for $Y$: $Y_i(0, M_i(0))$ and $Y_i(1, M_i(1))$. Accordingly, we define $Y_i(d) \equiv Y_i(d, M_i(d))$, so that the joint counterfactual need not be written explicitly when the mediator takes its natural value under treatment $d$. Thus we can simply write the total effect (TE) as
\begin{align}
    TE \equiv \E[Y_i(1) - Y_i(0)].
\end{align} 

When it is necessary to consider mediator values other than $M_i(d)$, we explicitly write both arguments as $Y_i(d, m)$. This is needed only where we describe mediation-related estimands to contrast them with our approach (see Section~\ref{subsec:compareestimands}, Table~\ref{tab:compare-estimands}).

\subsection{Defining the TRACE}

Our target quantity of interest is a total effect (TE) of $D$ on $Y$, averaged over only units that, if treated, would have $M=1$.\footnote{To simplify notation, we omit the subscript $i$ in the text for the remainder of the paper.} 
\begin{align}
    \text{TRACE} &\equiv \E[Y_i(1) - Y_i(0) \mid M_i(1)=1] \\
    &= TE[M_i(1)=1]
\end{align}

We also define the TE among the units that, if treated, would have $M=0$, which we call TRACE(0): 
\begin{align}
    \text{TRACE(0)} &\equiv \E[Y_i(1) - Y_i(0) \mid M_i(1)=0] \\
    &= TE[M_i(1)=0]
\end{align}

\subsection{Partial identification of the TRACE}

Identification of the TRACE is complicated by $M$'s status as a post-treatment variable, combined with possible unobserved common cause confounders of $M$ and $Y$. If investigators believed all common causes of $M$ and $Y$ were observed (i.e., there was no $U$ on Figure~\ref{fig:trace-dag}), it would be possible to identify the effect of $D$ on $Y$ conditioning on $M$, because all the paths opened by this conditioning could be closed again.\footnote{Representing the TRACE on a graphical causal model similar to Figure~\ref{fig:trace-dag} is possible but complicated, as it requires modifications to show $M(d)$ on the graph; see~\ref{app.dag}.} However, we consider this unrealistic in many settings, and of no interest for the problems on which we focus. For example, if $D$ is a community-driven development program, and $M$ is an indicator of whether a project chosen by the community was actually built, one expects many unobservable features to be common causes both of project implementation and the outcomes of interest $Y$. We therefore focus on the case in which such $M-Y$ confounding cannot be ruled out.

\subsubsection{No-assumption trimming bounds}

We begin by establishing bounds on the TRACE that rely only on assumptions standard in a randomized experimental setting. The first component of the TRACE,
$\E[Y(1) \mid M(1)=1]$, is directly identifiable. Among treated units, the observed outcome equals $Y(1)$ and those with $M=1$ are precisely those with $M(1)=1$, both by consistency. Hence we can identify
$\E[Y(1) \mid M(1)=1, D=1]$, which under ignorability of $D$ yields
$\E[Y(1) \mid M(1)=1]$. We denote the corresponding sample estimator as
$\overline{Y(1)}_{M(1)=1}$.

The second component,
$\E[Y(0) \mid M(1)=1]$, presents the difficulty. To observe $Y(0)$ we must examine units with $D=0$, but among these units we do not observe $M(1)$ and therefore cannot %determine which units would have
identify units with $M(1)=1$. Nevertheless, we can identify and estimate $\Pr(M(1)=1)$ as $\widehat{\Pr}(M=1 \mid D=1)$. This allows us to place sharp bounds on $\E[Y(0) \mid M(1)=1]$. The lowest (highest) possible value of this mean, $\overline{Y(0)}_{\text{low}}$ ($\overline{Y(0)}_{\text{high}}$), is the average over the lowest (highest) $\widehat{\Pr}(M=1 \mid D=1)$ fraction of the control outcome distribution. These bounds imply corresponding bounds on the TRACE estimate in a sample:

\begin{align}
\widehat{\text{TRACE}}_{\text{low}}
&=
\overline{Y(1)}_{M(1)=1}
-
\overline{Y(0)}_{\text{high}}, \\
\widehat{\text{TRACE}}_{\text{high}}
&=
\overline{Y(1)}_{M(1)=1}
-
\overline{Y(0)}_{\text{low}}.
\end{align}

This bound is sharp: values as low as $\text{TRACE}_{\text{low}}$ and as high as $\text{TRACE}_{\text{high}}$ (inclusive) can be obtained with the observed data.\footnote{A similar approach has been used to derive sharp bounds for the survivor average causal effect under monotonicity \citep{lee2009training, imai_sharp_2008, zhang_estimation_2003, horowitz1995identification} or ``conditional monotonicity'' \citep{Samii_Wang_Zhou_2025, semenova2020generalized}. We briefly describe how to incorporate a monotonicity assumption in Section \ref{data-driven-bounds}.}

\subsubsection{Leveraging assumptions on TRACE(0)} 
While the no-assumption bounds provide a valuable starting point, they often will not be independently informative. The core of our approach focuses on an additional source of leverage investigators may have: defensible assumptions about how large the treatment effect could be for those who would have $M=0$ if treated. By the law of iterated expectations,
\begin{align}
TE &= \text{TRACE} \cdot Pr(M(1)=1) + \text{TRACE}(0) \cdot Pr(M(1)=0)  \\
\text{TRACE} &= \frac{TE - \text{TRACE(0)}\cdot Pr(M(1)=0)}{Pr(M(1)=1)},
\label{senseform}
\end{align}

\noindent where we suppress the $i$ index for quantities representing group proportions (e.g. $Pr(M(1)=1)$) to simplify notation. Given randomization of $D$, the TE is identifiable, as are $Pr(M(1)=1)$ and $Pr(M(1)=0)$:
\begin{align}
Pr(M(1)=m) &= Pr(M(1)=m \mid D=1) \quad \text{randomization of $D$} \\
&= Pr(M=m \mid D=1) \quad \text{consistency of $M(d)$}.
\end{align}

The only unknown on the right hand side of Expression~\ref{senseform} is TRACE(0). Thus, for any postulated range of TRACE(0) values, we can recover the implied range of TRACE values. The resulting range will be narrower when $Pr(M(1)=0)$ is smaller, i.e. when the fraction of ``implementers" is higher. Additionally, the entire procedure can be done conditioning on values of $X$.

\subsubsection{Estimation and inference}

For estimation, we use sample analogs for each element in Equation~\ref{senseform}: $\widehat{TE}=\overline{Y}_{D=1}-\overline{Y}_{D=0}$, and $\widehat{Pr}(M(1)=m)=\widehat{Pr}(M=m \mid D=1)$. When conditioning on covariates $X$ is required,  $\widehat{TE}$ can instead be estimated via a suitable model. Postulated $\text{TRACE}(0)$ values are then plugged in to produce corresponding TRACE values. For inference, we employ the percentile bootstrap: resampling the data with replacement, re-estimating the $\text{TRACE}$ in each replicate (holding $\text{TRACE}(0)$ fixed at its postulated value), and forming 95\% confidence intervals using the 2.5th and 97.5th percentiles of the resulting empirical distribution.

\subsubsection{Anticipated types of assumptions on TRACE(0)} 
We emphasize that the investigator \textit{is not expected to know, nor postulate a single value of TRACE(0).} On the contrary, maintaining conservative uncertainty over this quantity is central to the approach's credibility. Nevertheless, inferences can sometimes be made for the price of defensible claims regarding the reasoned bounds on this quantity. Our experience to date suggests several common forms of assumptions, though others are possible.
\begin{itemize}

\item \textit{Arguments that $TRACE(0) = 0$.} In some cases we have reason to believe that, absent the ``activation'' of $M$, there can be no effect of $D$ on $Y$. For example, in the police violence application, if the police do not make a stop ($M=0$), there cannot be police violence during that stop. Together with an assumption of no defiers (or no effect among defiers), this would support the assumption of $\text{TRACE}(0)=0$. This assumption is distinct from the exclusion restriction as it allows a direct effect. We elaborate on this distinction in Section~\ref{subsec:compareestimands}.

\item \textit{Arguments that $TRACE(0) \approx 0$.} In other circumstances we cannot argue that an effect is precisely impossible when $M=0$, but are equipped to argue there should be, at most, a very small effect. For example, if the treated are enrolled in an exercise program but never participate, it would be unlikely to have a substantial effect on their health. However,  we cannot rule out some small effect, e.g. through triggering other behavioral changes.

\item \textit{Arguments that $|TRACE(0)| < |TRACE|$}. In many other cases, we cannot be certain about the impact of $D$ on $Y$ in units with $M(1)=0$, but we can be convinced that this effect is less than the effect among those with $M(1)=1$. This means that, while we do not know the TRACE, we may be willing to argue that TRACE(0) is smaller (in absolute value) than it. For example, in the community policing application below, we argue that a community randomly assigned to receive the community policing intervention but not showing evidence of its implementation will not have as large an effect as in those places where we would see evidence that the program occurred as intended.

\item \textit{Arguments that TRACE(0) is opposite in sign to the TRACE}. In some cases investigators may argue that those with $M(1)=0$ show an effect opposite in direction to that in the  $M(1)=1$ group. For example, if a development program in a village promises to provide something ($D=1$) but this fails to materialize ($M=0$), it may engender greater mistrust or anger. 

\end{itemize}

Investigators may argue for one of these assumptions or others to arrive at a defensible range of TRACE results. Alternatively, this approach allows users to transparently show what values of TRACE(0) are required to reach a particular conclusion about the TRACE (e.g. a significant effect in the intended direction among implementers). The researcher (and reader) can then reason about plausibility and defensibility of these assumed values.

\subsection{Sharpness of combined bounds}

The approach we recommend and illustrate below presents both the no-assumption trimming bounds and the partial identification results obtained for any postulated range of TRACE(0), highlighting the region consistent with both. This ensures that assumptions about TRACE(0) can provide identification leverage where possible, while results remain sharp by simultaneously enforcing the no-assumption trimming bounds.\footnote{Because we obtain closed-form bounds under the assumptions considered here, computational approaches such as AutoBounds \citep{duarte2024automated} are unnecessary for the present analysis. However, AutoBounds could be valuable in cases where researchers wish to impose additional assumptions or alternative constraints for which analytic bounds are difficult to derive, presuming also that the outcome $Y$ can reasonably be discretized. \ref{app:autobounds} briefly outlines how the TRACE estimand can be polynomialized to support such an approach. 
}

To see that the TRACE(0)-based bounds are sharp when they do not violate the trimming bounds, note first that TRACE values are algebraically determined by (i) the postulated value of TRACE(0) and (ii) two quantities identified from the data ($TE$ and $\Pr(M(1)=1)$). TRACE values generated in this way are thus compatible with the data, as long as the assumed TRACE(0) is not itself inconsistent with the data. This condition could be guaranteed by imposing trimming bounds on TRACE(0), analogous to those derived for the TRACE above. However, this is unnecessary because these TRACE(0) trimming bounds are precisely those that imply the TRACE trimming bounds. This follows directly from the algebraic relationship linking the TRACE and TRACE(0), a result which we verify 
in~\ref{app:samebounds}. Consequently, imposing the no-assumption trimming bounds on the TRACE and combining them with bounds obtained under assumptions on TRACE(0) yields a sharp partial identification region.

\subsection{Related estimation approaches}

Several common estimation strategies are worth noting for their widespread use, regardless of whether they target a well-defined causal estimand in this setting. As already noted, one of these is post-treatment conditional estimation: comparing mean outcomes for treated and untreated units with observed M = 1. We compare this analytically to the TRACE in Section~\ref{subsec:compareestimands} below. However, several other practices merit comment.   
\begin{itemize}
\item The intention-to-treat (ITT) approach evaluates causal effects based on the assigned treatment when not all participants adhere to treatment assignment. Our proposal essentially formalizes the intuition that the ITT may be  ``diluted,'' then explores the ``un-diluted'' values implied by varying assumptions on TRACE(0). 
\item ``As-treated'' analysis estimates the effect of the realized treatment rather than the assigned treatment, reclassifying units according to what they actually received regardless of assignment. This estimate will suffer bias from confounding if characteristics that moved a participant to treatment implementation are associated with characteristics that affect a participant's outcome. 
\item ``Per-protocol'' analyses instead retain the original treatment arms but restrict each to participants who adhered to their assigned protocol---retaining those with $M = 1$ in the treated group and $M = 0$ in the control group. This conditions on a post-treatment variable in both arms, breaking the randomization. 
\end{itemize}

Finally, a related practical approach is to model $M(1)$ for all units based on available (pre-treatment) covariates. The estimated $\widehat{M}(1)$ can be conditioned on (see e.g. \citealp{vanacore2024effect}). This characterizes the total effect across groups defined by their pre-treatment covariates. The predicted $\widehat{M}(1)$ is at best an imperfect proxy for $M(1)$; nevertheless, this approach may prove valuable when it is less important to precisely learn the TRACE but desirable to have some directional estimate of how it changes as a function of relevant pre-treatment covariates. 

\subsection{Comparison to related estimands}\label{subsec:compareestimands}

Several existing estimands address queries related to post-treatment variables, but differ from the TRACE in the groups they include and the identification opportunities they present. In making these comparisons, it is helpful to follow the structure of the principal stratification framework \citep{frangakis_principal}. Specifically, imagine that all units fall into one of four possible strata: compliers ($M(1) = 1, M(0) = 0$), defiers ($M(1) = 0, M(0) = 1$), always-takers ($M(1) = 1, M(0) = 1$), and never-takers ($M(1) = 0, M(0) = 0$). The TRACE is the average treatment effect for always-takers and compliers, while TRACE(0) is the average effect over never-takers and (if present) defiers.

Table~\ref{tab:compare-estimands} compares related estimands. Though ours is unique (excepting equivalences in special cases), the primary emphasis of our contribution lies in the relatively straightforward partial identification strategy it allows.

\begin{table}[]
 \renewcommand{\arraystretch}{1.5} % Default value: 1
\centering
\begin{tabular}{@{\extracolsep{0pt}} p{0.1\linewidth} p{0.35\linewidth} p{0.12\linewidth} p{0.4\linewidth} }
\hline 
\hline  
Estimand & Definition &  Among Whom & Identification Assumptions\\
\hline 
LATE (IV) & $\E[Y(1) - Y(0) \mid M(0) = 0, M(1)=1]$ &  compliers &  relevance, exogeneity, exclusion restriction\\
SACE & $\E[Y(1) - Y(0) \mid M(0) = M(1) = 1]$ & always-takers & monotonicity,   $M(1) \indep M(0)$,\; $\E[Y(1) \mid \text{AT}] = \E[Y(1) \mid \text{C}]$, \; $\E[Y(0) \mid \text{AT} ] = \E[Y(0) \mid \text{D}]$ given $X$\\
   
CDE(m) & $\E[Y(D=1, M=m) - Y(D=0, M=m)]$ & all units & unconfoundedness of $M-Y$ \\         
TRACE  & $\E[Y(1) - Y(0) \mid M(1) = 1]$ & compliers,  always-takers  & postulated TRACE(0)\\
\hline  
\end{tabular}
\caption{Comparison of different estimands for settings in which post-treatment variables are relevant. AT: always-taker; C: complier.}
\label{tab:compare-estimands}
\end{table}

\paragraph{Comparison to instrumental variables.} The Local Average Treatment Effect (LATE), also called the Complier Average
Causal Effect, \citep{imbens_identification_1994,baker_paired_1994} is
\begin{equation}\label{eq.late}
  \E[Y_i(1) - Y_i(0) \mid M_i(0) = 0, M_i(1) = 1].
\end{equation}

This estimand differs from the TRACE in two primary ways.\footnote{
As noted briefly above, the IV causal structure relevant here would consider our $M$ the treatment, and our $D$ the instrument for it. Relatedly, the $Y_i(\cdot)$ in Expression~\ref{eq.late} refers to $Y_i(m)$ in the IV setting, and the causal effect among compliers is thus the effect of $M$ on $Y$, not the effect of $D$ on $Y$ we are investigating. That said, for compliers 
$D$ and $M$ are required to be the same, so the notational difference is immaterial in this particular case.} First, they differ definitionally in that the LATE considers the average effect among compliers, while the TRACE considers the average effect among compliers and always-takers. These estimands become equivalent when we can rule out always-takers (through one-way non-compliance), or when effects for compliers and always-takers are the same. Second, the LATE is the target estimand for the IV approach, which for identification requires that $D$ only affects the outcome through $M$, i.e.\ that there is no direct effect of $D$ on $Y$. The $D \to Y$ path violates this core IV assumption, and is an important contributor to the total effect we are seeking to learn (see Figure~\ref{fig:trace-dag}).\footnote{For such settings, partial identification strategies have been developed for bounded potential outcomes \citep{manski1990nonparametric, manski1998monotone}.}

Consider, for example, cases where the investigator argues that TRACE(0) = 0. This may occur in ``Type 3'' settings where it is impossible for $D$ to affect $Y$ when $M = 0$, and we can further argue either for monotonicity (no defiers) or for no effect among defiers. The assumption of ``no effect of $D$ on $Y$ when $M = 0$'' may sound similar to the exclusion restriction, but they differ in a key way: the exclusion restriction requires the total absence of a direct $D \to Y$ effect, while TRACE(0) = 0 requires no direct effect only for those with $M(1) = 0$ (never-takers and, if present, defiers). In some cases, a direct $D \to Y$ effect in the ``reactive'' subgroup is not only expected, but is precisely our quantity of interest. In the police violence setting discussed below, for example, we wish to study---not rule out---a possible effect of police-perceived race ($D$) on the use of violence ($Y$) when a stop occurs ($M = 1$).

\paragraph{Comparison to survivor average causal effect.} The ``survivor average causal effect'' (SACE) arises in settings where the outcome is censored due to a post-treatment factor, such as survival \citep{tchetgen_identification_2014, egleston_estimation_2009, imai_sharp_2008, hayden_estimator_2005, zhang_estimation_2003, rubin_causal_2000, robins_new_1986} or sample selection/attrition \citep{semenova2020generalized, lee2009training}.  The SACE is defined as
\begin{equation}
\text{SACE} = \E[Y_i(1) - Y_i(0) \mid M_i(0) = 1, M_i(1) = 1].
\end{equation}

It considers only the effect among always-takers, i.e.\ ``always-survivors'' when $M = 1$ is survival to endline. Because we cannot observe which units are always-takers, point identification strategies rely on assumptions that ensure mean values of $Y_i(1)$ or $Y_i(0)$ do not differ between certain principal strata.

The TRACE differs from the SACE only by including compliers. The survival settings in which the SACE originates can introduce complications for our approach by rendering the total effect unidentified. As discussed in Section~\ref{subsec.app1}, this can sometimes be addressed, particularly in settings where $M$ is a ``necessary condition mediator'' and $D$ is assumed to be monotonic in $M$. Outside of the survival setting, the SACE is relevant when reinterpreted as the average effect among always-takers, i.e.\ where $M$ does not affect loss to follow-up (see e.g. \citealp{hudgens_causal_2006}). Existing work on partial identification and trimming bounds for the SACE has proposed similar bounds for $\E[Y(d) \mid AT]$ \citep{Samii_Wang_Zhou_2025, semenova2020generalized, lee2009training, imai_sharp_2008, zhang_estimation_2003, horowitz1995identification}. We discuss similar bounds for the TRACE in Section~\ref{data-driven-bounds}.

\paragraph{Comparison to mediation analysis.} A variety of mediation approaches decompose the total effect into the portion attributable to direct action of the treatment and the portion operating indirectly through changes in the mediator (e.g. \citealp{robins1992identifiability, pearl_direct_2001, imai_general_2010, imai_comment_2014,  pearl_interpretation_2014, acharya_explaining_2016}). Our causal query is of a different kind: rather than partitioning pathways of an effect defined over a common population, we seek a
\textit{total} effect for a subgroup defined by the potential value of the mediator. Because these approaches are organized around different goals, a formal comparison of the underlying quantities offers little improvement in understanding over simply recognizing this distinction. 

\paragraph{Comparison to post-treatment conditional estimation.} While the problems with estimating effects conditional on $M = 1$ are well known (see e.g. \citealp{montgomery2018conditioning}), comparing this approach to the TRACE can be instructive. Defining $\text{DIM}_{M=1} \equiv \E[Y \mid M = 1, D = 1] - \E[Y \mid M = 1, D = 0]$, we can show that under monotonicity of $M$ in $D$,\footnote{See~\ref{app:naivecomp}. An intuition for this difference is that the TRACE and $\text{DIM}_{\text{M=1}}$ obtain $Y(1)$ values from the same group of treated units (always-takers and compliers). But they differ in the untreated comparison group they employ: the TRACE is interested in $Y(0)$ among always-takers and compliers (to match the treated), whereas the $\text{DIM}_{\text{M=1}}$ only uses $Y(0)$ values from always-takers. Accordingly the two are separated by $\E[Y(0)\mid AT] - \E[Y(0) \mid C]$, rescaled by the proportion of the TRACE subgroup that is made up of compliers.}
  \begin{equation}
  \text{TRACE} - \text{DIM}_{M=1} = \left(\E[Y_i(0) \mid AT] - \E[Y_i(0) \mid C]\right) \left(\frac{\Pr(C)}{\Pr(AT) + \Pr(C)}\right).
  \end{equation}

 \noindent In ``Type 3'' cases (where $M = 0$ guarantees $Y = 0$), 
  $Y(0) = 0$ among compliers, reducing this to
  \begin{equation}\label{eq:dimm1comp}
  \text{TRACE} - \text{DIM}_{M=1} = \E[Y_i(0) \mid AT] \left(\frac{\Pr(C)}{\Pr(AT) + \Pr(C)}\right),
  \end{equation}
  which for non-negative outcomes implies TRACE $\geq$ DIM$_{M=1}$.

\subsection{Monotonicity-based trimming bounds}
\label{data-driven-bounds}

A potentially tighter trimming bound is possible when (i) $M$ is observed for all units and (ii) the user is willing to assume monotonicity ($M(1) \geq M(0)$, i.e., no defiers). The problematic quantity for identification, 
$\E[Y(0) \mid M(1)=1]$, then consists of only always-takers (AT) and compliers (C). Specifically, letting $\alpha \equiv \Pr(AT \mid M(1)=1)$,
\begin{align}
\E[Y_i(0)\mid M_i(1)=1]
&= \E[Y_i(0)\mid AT] \alpha + \E[Y_i(0)\mid C] (1-\alpha).
\label{eq:mix}
\end{align}

\noindent Under monotonicity, untreated units with $M=1$ are necessarily always-takers, so
$\E[Y(0)\mid AT]$ is identified as $\E[Y \mid D=0, M=1]$. Further, $\alpha$ is identified as $\frac{\Pr(M=1 \mid D=0)}{\Pr(M=1 \mid D=1)}$, 
with sample estimator $\hat{\alpha}$. The unidentified portion of $\E[Y(0)\mid M(1)=1]$ is solely due to $\E[Y(0)\mid C]$, which we can bound based on the distribution of outcomes for units with $D=0$ and $M=0$. This group includes compliers and never-takers (NT), with an identified share $\pi$ being compliers: \[
\pi \equiv 
\frac{\Pr(C)}{\Pr(C)+\Pr(NT)} = \frac{\Pr(M=1\mid D=1)-\Pr(M=1\mid D=0)}{\Pr(M=1\mid D=1)-\Pr(M=1\mid D=0) + \Pr(M=0\mid D=1) }
\]
We can then construct trimming bounds  $\widehat{\E}[Y(0)\mid C]_{\text{low}}$
and
$\widehat{\E}[Y(0)\mid C]_{\text{high}}$
as the means of the lowest and highest $\hat{\pi}$ fractions of
outcomes in this group. These yield monotonicity-based trimming (MT) bounds on $\widehat{\E}[Y(0) \mid M(1)=1]$ of
\begin{align}
\widehat{\E}[Y_i(0) \mid M_i(1)=1]_{\text{low}}^{\text{MT}}
&=
\widehat{\E}[Y_i(0)\mid AT]\hat{\alpha}
+
\widehat{\E}[Y_i(0)\mid C]_{\text{low}}(1-\hat{\alpha}), \\
\widehat{\E}[Y_i(0) \mid M_i(1)=1]_{\text{high}}^{\text{MT}}
&=
\widehat{\E}[Y_i(0)\mid AT]\hat{\alpha}
+
\widehat{\E}[Y_i(0)\mid C]_{\text{high}}(1-\hat{\alpha}).
\end{align}

\noindent which finally yields the monotonicity-based trimming bounds on TRACE,
\begin{align}
\widehat{\text{TRACE}}_{\text{low}}^{\text{MT}}
&=
\overline{Y(1)}_{M(1)=1}
-
\widehat{\E}[Y_i(0) \mid M_i(1)=1]_{\text{high}}^{\text{MT}}, \\
\widehat{\text{TRACE}}_{\text{high}}^{\text{MT}}
&=
\overline{Y(1)}_{M(1)=1}
-
\widehat{\E}[Y_i(0) \mid M_i(1)=1]_{\text{low}}^{\text{MT}}.
\end{align}

\section{Examples}

\subsection{Hypothetical demonstration: Perceived race, police stops, and violence}\label{subsec.app1}
Here, we demonstrate the use of this approach for understanding the influence of race on police behavior, a setting in which selection into administrative records data has made causal inference challenging. Consider a scenario in which police officers observe drivers and have a perception (accurate or not) of each driver's race. Let $D=1$ indicate that the police perceived the driver to be from a minority group.  The police may then choose to stop the vehicle ($M$). During that stop, police violence may or may not occur ($Y$). Only encounters that result in stops appear in police administrative data. Thus, analyses using only these data automatically condition on $M=1$, leading to post-treatment bias when comparing rates of police violence by driver race among recorded stops \citep{knox_administrative_2020}.\footnote{This setting is not experimental, so in an actual empirical study of this question, unobserved confounding would pose additional problems beyond the identification concerns related to post-treatment conditioning discussed here and elsewhere.}

Because $Y$ is defined specifically as ``violence during a police stop'', $Y=0$ when no stop occurs ($M=0$) by construction, making this an example of a ``necessary condition mediator'' (Type 3). Recall that TRACE(0) is the total effect among never-takers (units with $\{M(1)=0,M(0)=0\})$, and defiers (units with $\{M(1)=0, M(0)=1 \}$). Suppose we can rule out defiers, arguing that if a person is not stopped when perceived to be a minority, they would not be stopped when perceived to be non-minority. Consequently, the $M(1)=0$ group would include only never-takers. Because this group is never stopped regardless of perceived race, it follows that $Y(1)=Y(0)=0$. Thus, the total effect among never-takers is deterministically zero, implying 
$TRACE(0)=0$, and point identifying the TRACE.\footnote{Even when we cannot rule out defiers, we may still be able to argue that $\text{TRACE}(0)$ is nearly zero, which may inform inference.  Consider cases in which a police officer would stop an individual only if they were perceived as non-minority, e.g. during a manhunt for a white suspect. Information about the nature of these stops and how often such manhunts 
occur could put reasonable limits on the ratio of never-takers to defiers. If the proportion of defiers is expected to be much smaller than the proportion of never-takers, $TRACE(0)$ will be very close to 0, even if the effect among defiers is not. Partial identification can then proceed with $TRACE(0)$ values near $0$. We thank an anonymous reviewer for raising this consideration.}

Estimation requires only the sample moments needed to estimate (i) the total effect, and (ii) the proportion with $M(1)=1$. The former requires knowing the share of encounters with perceived-minority drivers that result in violence ($Pr(Y=1 \mid D=1)$), as well as the corresponding share for perceived-non-minority drivers ($Pr(Y=1 \mid D=0)$). This may not be directly available, but could be backed out from alternative combinations of information, such as the fraction of cases with violence involving perceived minority drivers ($Pr(D=d \mid Y=1)$), the overall proportion of police-perceived minority drivers ($Pr(D=1)$) in the relevant population, and the overall proportion of encounters resulting in violence ($Pr(Y=1)$). However, estimating the second quantity, $Pr(M(1) = 1)$, requires data on the total number of encounters with perceived minority drivers, in addition to data on the number of minority stops. \citet{knox_administrative_2020} similarly conclude that the total effect can be point identified in this setting given data on the total number of perceived minority and non-minority encounters.

Finally, investigators can directly estimate  $\text{DIM}_{\text{M=1}}=\E[Y \mid D = 1, M = 1]-\E[Y \mid D = 0, M = 1]$, using only data on encounters that result in stops. As noted in Section~\ref{subsec:compareestimands}, under monotonicity and with a ``necessary condition mediator,'' the TRACE will exceed this quantity by $\E[Y(0) \mid AT]\left(\frac{Pr(\text{C})}{Pr(\text{C})+Pr(\text{AT})}\right)$. Note also that $\E[Y(0)\mid AT] = \E[Y \mid D=0, M=1]$, which is also identifiable using only data from stops. Since this term is non-negative, $\text{TRACE} \geq 
\text{DIM}_{\text{M=1}}$. Further, when $\frac{Pr(\text{C})}{Pr(\text{C})+Pr(\text{AT})}$ is unknown, the TRACE remains bounded between $\text{DIM}_{\text{M=1}}$ and $\text{DIM}_{\text{M=1}} + \E[Y(0) \mid AT]$.\footnote{\ref{app:klmcomp} applies this to values available from \cite{knox_administrative_2020}.}

\subsection{Effects of community policing on mob violence in Liberia}

\paragraph{Setting.} Our first empirical application is an experimental study by \citet{morse_strengthening_2024} examining the effects of a community policing intervention in Monrovia, Liberia on mob violence, overall instances of crime, perceptions of security, and crime reporting.\footnote{This study was part of a coordinated multi-site trial across six countries \citep{blair_community_2021}.} The intervention, implemented in collaboration with the Liberian National Police (LNP), involved community town hall meetings, increased foot patrols, and the formation and training of local security groups known as Community Watch Forums. Forty-five out of 93 eligible communities were randomly selected to receive the program over a period of 10 months. Outcomes were measured primarily using community surveys administered three months after the end of the intervention. 
\citet{morse_strengthening_2024} finds that the intervention meaningfully reduced instances of mob violence, but did not affect overall crime, perceptions of security, or crime reporting.

The intervention incorporated training and capacity building for Community Watch Forums (CWFs), groups of citizens who cooperate  directly with the police. \cite{morse_strengthening_2024} attributes the intervention's success at reducing mob violence to this component in  particular. We use (community awareness of) the presence of a CWF as our primary mediator of interest $M$, setting $M=1$ if more than 20\% of community survey respondents at endline reported the existence of a CWF. Our main outcome of interest $Y$ is the average number of mob violence instances reported by community survey respondents in the past year. The TRACE represents the effect of the community policing intervention on reported mob violence \emph{in communities where assignment to the intervention would have resulted in the formation of a successful CWF} (hereafter ``implementing types").\footnote{Community security groups existed in some form in many areas prior to the intervention. Respondents may also believe there is a CWF, even if one was not created as a result of this intervention. As a result, it is possible for respondents in control communities to report the presence of a CWF.}

\begin{figure}
    \centering    \includegraphics[width=0.7\linewidth]{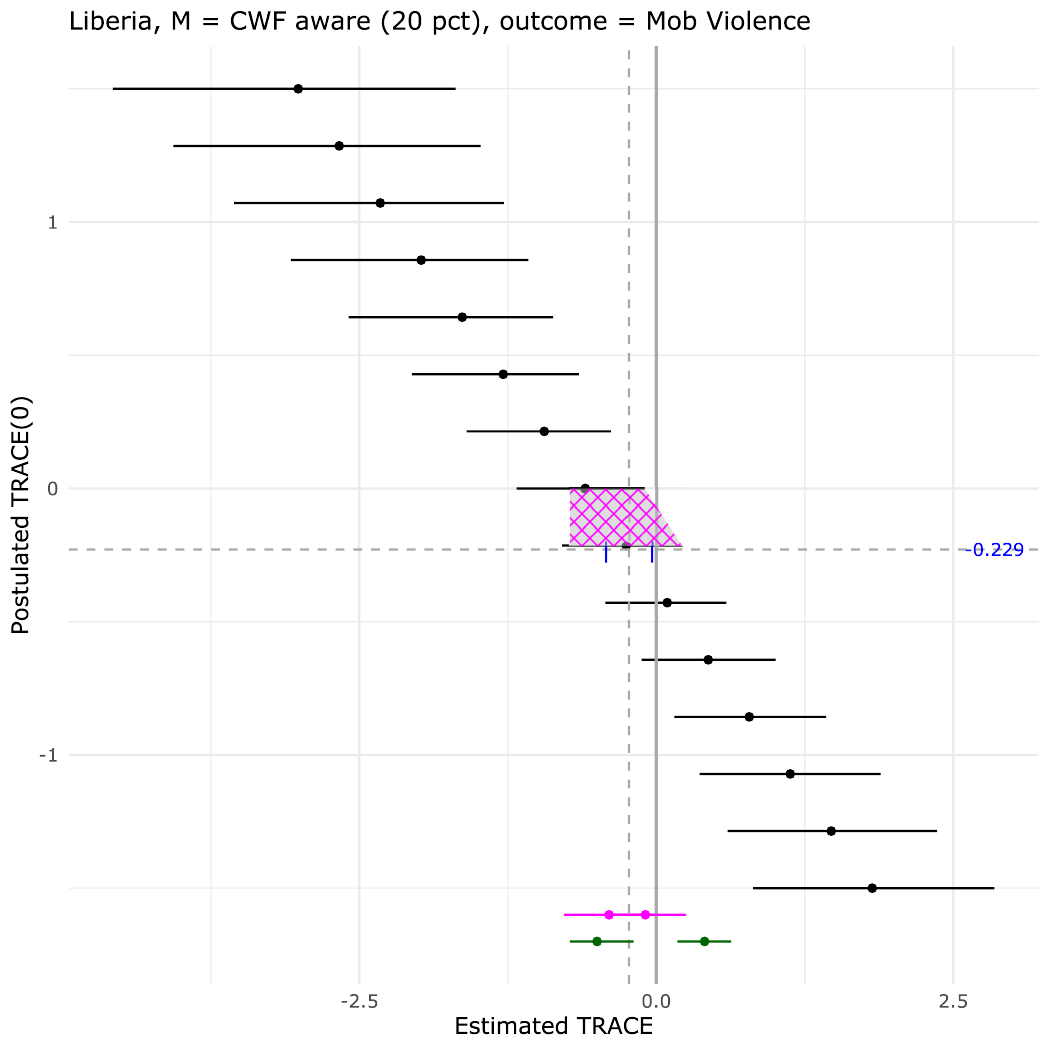}
    \caption{Estimated TRACE of community policing intervention on reported mob violence, among communities for which the intervention would have produced a successful Community Watch Forum (implementing types), across postulated values of the treatment effect among non-implementing types (TRACE(0)). Line segments represent 95\% confidence intervals. The number on the far right represents the ``naive" ITT estimate, assuming constant treatment effects across strata of $M$. The grey shaded region represents the intersection of the no-assumption bounds and the range of estimates assuming a value for TRACE(0) less than or equal to the value of the TRACE, in the same direction. The green dots show the point estimates of the no assumption bounds, and corresponding 95\% confidence intervals obtained by bootstrap. The pink bounds represent the MT bounds described in Section \ref{data-driven-bounds}, and the pink crosshatch shows the intersection of the MT bounds with the assumptions on TRACE(0). All models include police zone (block) fixed effects and baseline levels of reported mob violence, averaged at the community-level, as a covariate.}
    \label{fig:morsecwfcontrols}
\end{figure}

\paragraph{Results.} Figure \ref{fig:morsecwfcontrols} illustrates what can be claimed about the effect among implementing types (TRACE) given any assumption about the effect among non-implementing types (TRACE(0)).\footnote{Replication materials sufficient to reproduce all results and figures are available in the replication archive located at: \url{https://doi.org/link}.} The vertical axis represents postulated values of TRACE(0), while the horizontal axis represents corresponding TRACE estimates. The no-assumption trimming bounds are shown in green, and correspond to point estimates of -0.438 and 0.397 for the lower and upper bounds on TRACE.

To illustrate interpretation, consider three possible assumptions. First, suppose we assume that the effect of the community policing intervention is the same among implementing and non-implementing types (i.e. that $\text{TRACE} = \text{TRACE}(0)$). Under this assumption, our point estimate of the TRACE is equivalent to that of the ITT, represented by the intersection of the dashed lines at an estimate of -0.23.\footnote{Our point estimate of the ITT (-0.23) differs slightly from the estimate of -0.28 in \cite{morse_strengthening_2024} for two reasons. First, while \cite{morse_strengthening_2024} analyzes outcome data at the individual level and clusters standard errors at the community, we measure the outcome at the community level, using the community-level mean, avoiding the clustering requirement. Second, while we use OLS, \cite{morse_strengthening_2024} uses weighted least squares, weighting by the product of inverse probability of community assignment to treatment and inverse probability of individual selection for the survey.} HC2-based confidence intervals for the ITT are shown using the (blue) vertical line segments around the point estimate. Note that confidence intervals for the TRACE are larger than for the ITT even where the two point estimates are equivalent. This is because the former represents an inference about a sub-group (those with $M(1)=1$).

Second, if we are willing to posit that the intervention had no effect on mob violence in non-implementing type communities (i.e. that $\text{TRACE}(0) = 0$), we recover an effect size among implementing types of -0.59, more than twice the magnitude of the ITT, in the hypothesized direction (a reduction in mob violence).

Third, and more practically, we would likely be unwilling to endorse either of the assumptions above, especially using an arbitrary cutoff for community awareness as our implementation indicator ($M$). However, we are more willing to entertain the assumption that the intervention produced a larger benefit (a more negative effect on mob violence) among implementing types than among non-implementing types. Researchers may argue this directly based on the necessity of $M$ to achieving a meaningful effect.\footnote{More rigorously, the assumption can be supported by reference to a combination of direct and indirect effects and their signs. To simplify, suppose we can rule out defiers, as is reasonable here. Then  $\text{TRACE}(0)$ is simply the effect among never-takers, which is a direct effect of $D$ on $Y$ in that group without the indirect effect through $M$. We might expect this to be small if $D$ is an intervention that cannot reasonably have a large (direct) effect in the absence of $M$. Meanwhile, the $\text{TRACE}$ is due to both always-takers and compliers. The always-takers again contribute only a direct effect, which may arguably be small. However, compliers can provide both a direct effect and the key indirect effect of interest, which might be much larger when $M$ is clearly important to achieve an effect.} If true, this implies  estimates for the TRACE that fall to the left of the vertical dotted line.\footnote{TRACE estimates to the right of the vertical dotted line imply a more beneficial (negative) effect of the community policing intervention on mob violence among non-implementing types than among implementing types.} If we wish to further assume that TRACE and $\text{TRACE}(0)$ have the same sign, we bound the TRACE on the left at the point where $\text{TRACE}(0)=0$. This range corresponds to point estimates of TRACE between -0.23 and -0.59. By combining the confidence intervals around these estimates with the confidence intervals around the no-assumption bounds, we obtain the valid region for TRACE (Figure~\ref{fig:morsecwfcontrols}, shaded region).  

The results also imply that any argument for a TRACE (point) estimate larger than -0.59 would require us to believe that the intervention produced an average \emph{increase} in mob violence in non-implementing type communities. This could be possible, for example, if police lacked the capacity to respond to increased demand for law enforcement resulting from other aspects of the intervention and -- in the absence of a viable lawful community-based alternative -- communities (would have) resorted to vigilante justice. Investigators and subject experts can reason about the plausibility of such opposite-directional effects among non-implementing types in this or any given setting.

\begin{figure}
\centering
\begin{subfigure}{.7\textwidth}
  \centering
  \includegraphics[width=.7\linewidth]{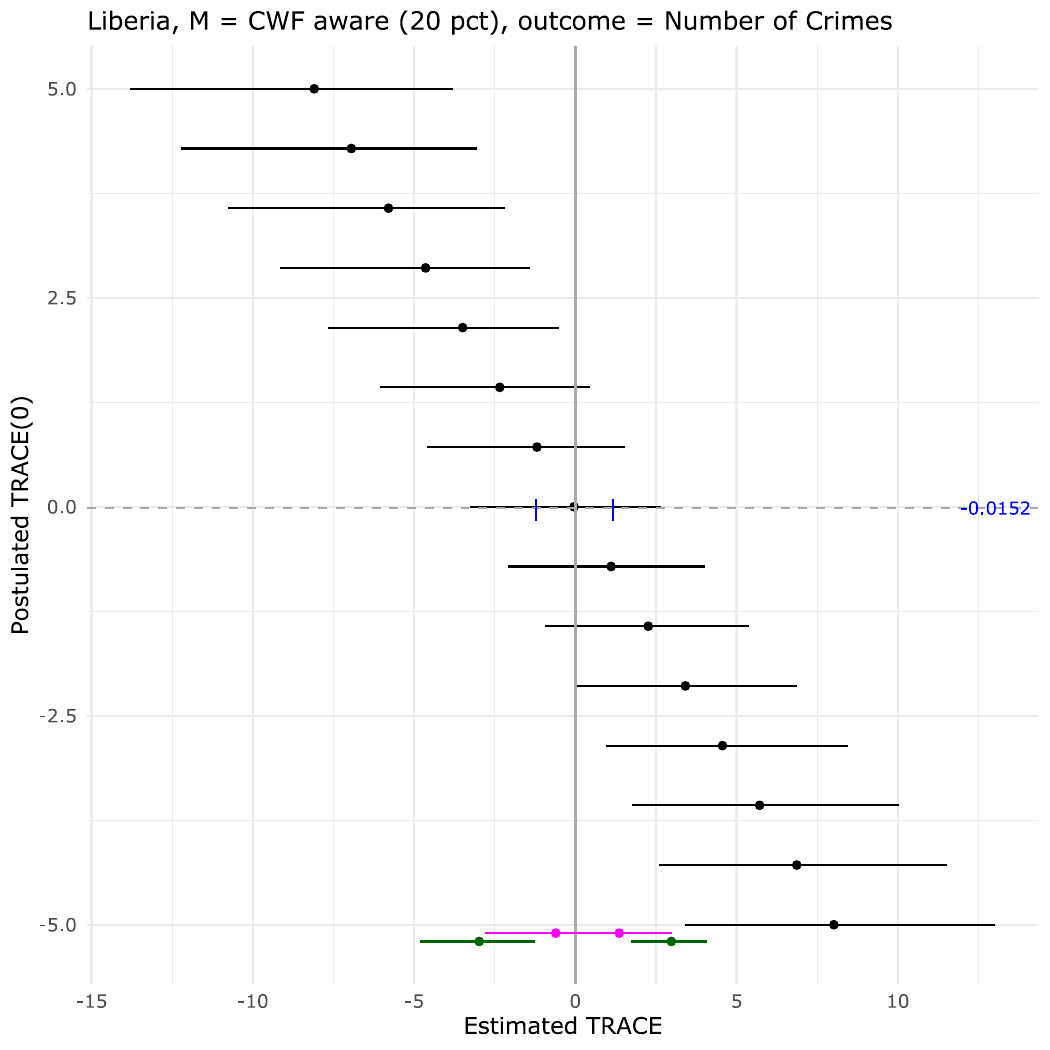}
  \caption{Community Watch Forum (CWF) awareness as the mediator ($M=1$ if 20\% of community survey respondents reported awareness of a CWF registered with police at endline).}
  \label{fig:morsecrimecwf}
\end{subfigure}\\
\vspace{0.5cm}
\begin{subfigure}{.7\textwidth}
  \centering
  \includegraphics[width=.7\linewidth]{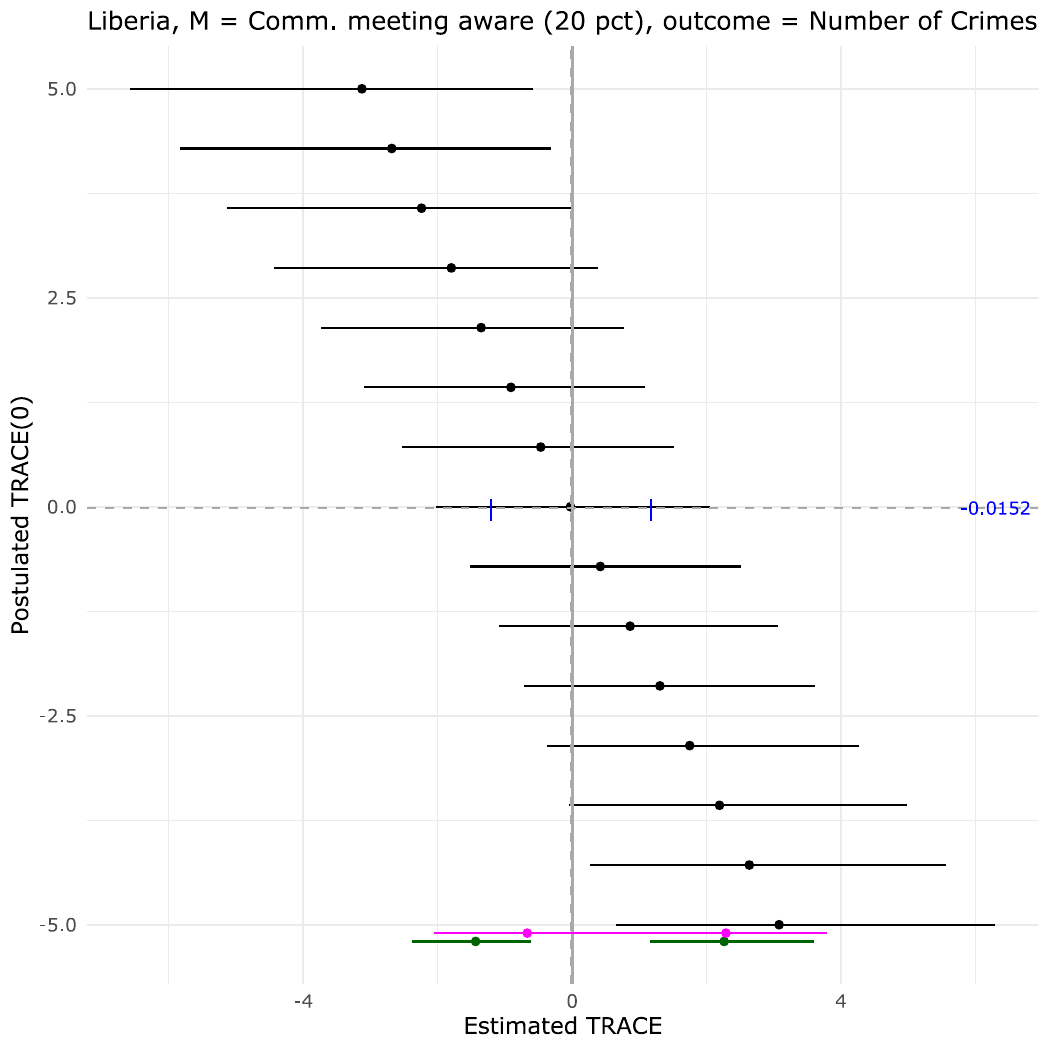}
  \caption{Community meeting attendance or awareness as the mediator ($M=1$ if 20\% of community survey respondents reported being aware of or having attended a community meeting with police at endline).}
  \label{fig:morsecrimemeeting}
\end{subfigure}
\caption{Estimated TRACE of community policing intervention on reported incidents of crime among implementing type-communities, across postulated values of the treatment effect among non-implementing-type communities(TRACE(0)), using two different implementation measures ($M$).
}
\label{fig:morsecrime}
\end{figure}

\paragraph{Assessing null findings.} Our approach can also be informative in the interpretation of null results. For example, unlike the encouraging findings regarding effects on mob violence, \citet{morse_strengthening_2024} does not find evidence that the community policing intervention reduced crime, improved security perceptions, or increased crime reporting. Given uneven implementation, we might wonder ``what would we have to believe about the effects of the intervention among non-implementing types to recover an effect in hypothesized direction among implementing types?'' 
Figure \ref{fig:morsecrime} answers this question when using crime incidence as the outcome.\footnote{For a list of crimes, survey respondents were asked whether they or anyone in their family were a victim of that type of crime in the past 6 months. The individual-level variable captures the total number of crimes reported across all categories. In our analysis, this variable is averaged at the community level.} In panel \ref{fig:morsecrimecwf}, we again use CWF-awareness for  $M$. In order to obtain a significant beneficial (negative) effect among implementing types, we would have to assume a harmful effect among non-implementing types larger in magnitude than the corresponding beneficial effect among implementing types. We expect this is  unlikely.

Panel \ref{fig:morsecrimemeeting} repeats this analysis replacing $M$ with an alternative implementation measure capturing another major component of the  intervention: community town hall meetings. Here, $M=1$ if more than 20\% of community survey respondents reported being aware of or having attended a community security meeting with police at endline. The results show that we cannot obtain a significant beneficial effect among implementing types (TRACE) even if the non-implementing group had a harmful effect (TRACE(0)) almost twice as large as the beneficial effect among implementers. This increases our confidence that the failure to find an effect on crime is not a result of uneven implementation across treated communities, at least insofar as we believe $M=1$ captures ``good enough'' implementation.\footnote{The monotonicity-based trimming bounds are also shown in Figures \ref{fig:morsecwfcontrols}, \ref{fig:morsecrimecwf}, and \ref{fig:morsecrimemeeting}. The``no defiers'' assumption here requires that no communities would have formed a police-registered CWF only if not assigned to the intervention. For the effect of the intervention on mob violence (Figure \ref{fig:morsecwfcontrols}), the MT bounds are (-0.40, -0.09).In this case, the MT bounds do not result in tighter bounds than the no-assumption bounds combined with reasoning about TRACE(0). In Figures \ref{fig:morsecrimecwf} and \ref{fig:morsecrimemeeting}, we see that the MT bounds remain wide, while reasonably defensible assumptions on TRACE(0) lead to more informative conclusions.}

\subsection{Empirical application 2: Effects of in-person canvassing on support for transgender rights}
\label{canvassing}

\paragraph{Setting.} Our second empirical application is a field experiment by \citet{broockman_durably_2016}, examining the effects of a door-to-door perspective taking intervention on attitudes toward transgender people and support for a transgender non-discrimination law. The intervention involved a ten-minute conversation between canvassers and voters in Miami, Florida, during which canvassers asked voters to talk about a time when they were judged for being different and encouraging them to see connections between their own experiences and those of transgender people. Registered voters who completed a baseline survey (n = 1,825) were randomly assigned to receive either the perspective taking intervention or a placebo intervention. Among the 501 voters who answered the door in either condition, outcomes were measured in follow-up online surveys three days, six weeks, and three months after the intervention.

\citet{broockman_durably_2016} find evidence that the intervention both reduced prejudice against transgender people and increased support for policies benefitting them: six weeks and at three months post-intervention, treated individuals were significantly more tolerant of transgender people and more supportive of an ordinance protecting them from discrimination in housing employment and public accommodations.\footnote{The authors did not find positive effects on support for the non-discrimination law during the first and second waves of outcome measurement. Beginning in the third wave (six weeks post-intervention), they revised the survey to define the term ``transgender." Positive effects on support are detected only after this point. We therefore focus our re-analysis on the third survey wave.}

Investigators may then wish to know what the effect of the intervention on policy attitudes would be if they could look just at individuals \emph{for whom the intervention (would have, if treated) produced a positive effect on attitudes toward transgender people}. For this application, we term this subset ``reactive types,'' noting that this shorthand is imperfect. We consider this an example of a ``mechanism of interest''  (type 2) question described above.
Our primary outcome $Y$ is support for the transgender non-discrimination law six weeks after treatment (wave 3), measured on a seven-point scale. Our mediator ($M$) captures the \emph{change} in subjective feelings toward transgender people from baseline. The authors measure attitudes toward transgender people during all waves using a standard 0-100 feeling thermometer, where a higher number represents ``warmer" feelings. We code $M=1$ if an individual's thermometer score increased between baseline and wave 3, and $M=0$ otherwise.\footnote{In the treatment group, thermometer scores increased between baseline and wave 3 for approximately 43\% of respondents.}

\begin{figure}
    \centering    \includegraphics[width=0.7\linewidth]{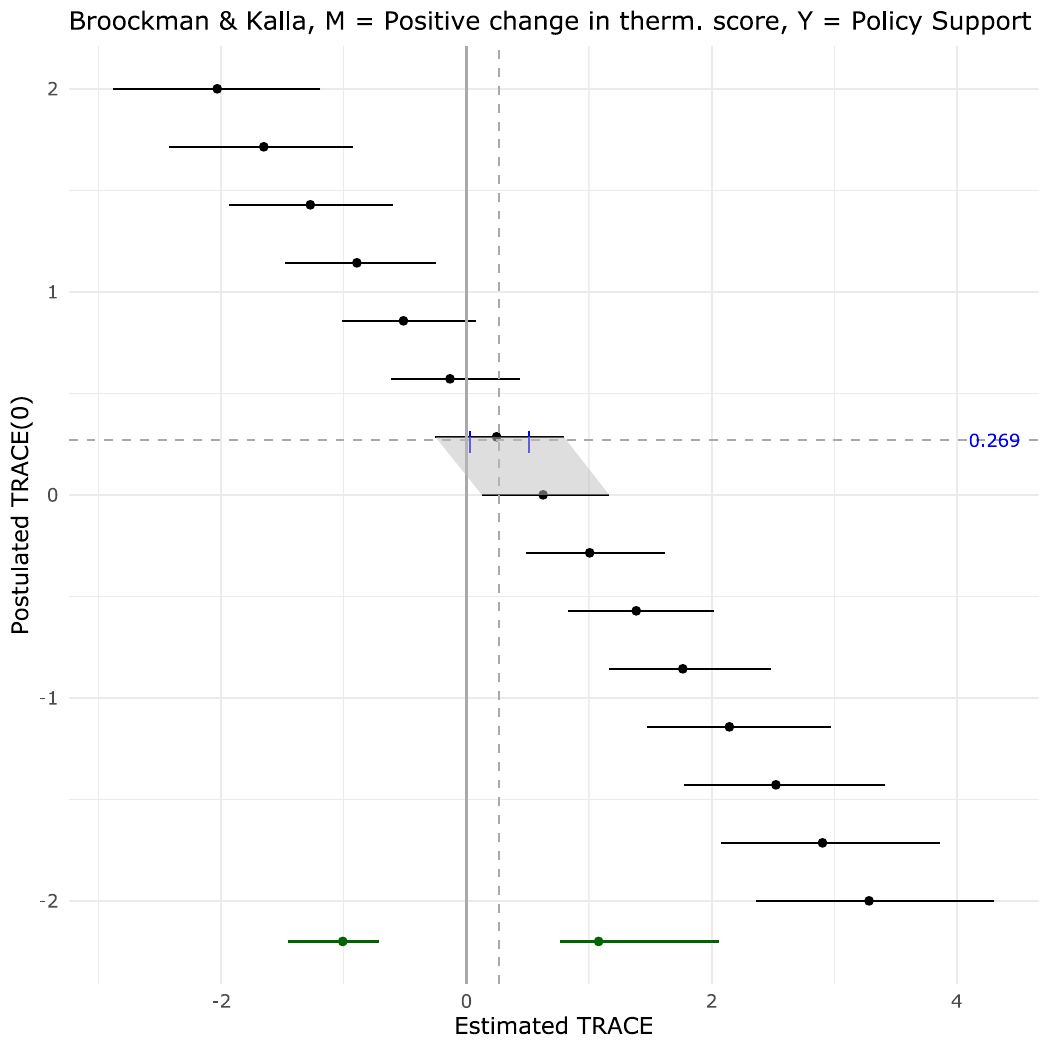}
    \caption{Estimated TRACE of in-person canvassing intervention on support for transgender non-discrimination law, among those for which the intervention would have led to an improvement in feelings toward transgender people from baseline (reactive types), across postulated values of the treatment effect among non-reactive types (TRACE(0)). Line segments represent 95\% confidence intervals. The number on the far right represents the ``naive" ITT estimate, assuming constant treatment effects across strata of $M$. The grey shaded region represents the intersection of the no-assumption bounds and the range of estimates assuming a value for TRACE(0) less than or equal to the value of the TRACE, in the same direction. The green dots show the point estimates of the no assumption bounds, and corresponding 95\% confidence intervals obtained by bootstrap. All models include control variables used in \citet{broockman_durably_2016}, including baseline levels of both transgender attitudes and policy support.}
    \label{fig:bkchange}
\end{figure}

\paragraph{Results.} Figure \ref{fig:bkchange} shows the results from this analysis. We illustrate interpretation with the same three assumptions employed in the previous application. First, if we assume the intervention was equally effective in increasing support for the non-discrimination law among individuals for whom it (would have) produced an improvement in subjective feelings towards transgender people and those for whom it would not have, our TRACE estimate is equal to the ITT (in this case, 0.27).\footnote{The original paper reports the Causal Average Complier Effect (CACE) rather than the ITT, adjusting for limited non-compliance with treatment assignment. While the CACE estimate reported in the paper (0.36) is larger than the ITT, the ITT estimate is still statistically significant.} Second, if we instead assume no effect of the intervention on policy support among non-reactive types, we recover an effect among reactive types of 0.63. Finally, if we merely wish to assume that the effect on policy support among reactive types is \emph{greater} than that among non-reactive types (but that the latter is still positively-signed), we obtain a range of estimates for the TRACE between 0.27 and 0.63. This range of estimates is fully contained in the interval suggested by the no-assumption trimming bounds, and is more informative than the no-assumption bounds alone. Any estimate greater than 0.63 would require an assumption of a negative (e.g. ``backlash'') effect among non-reactive types.\footnote{We do not attempt the alternative MT bounds here because we find it unlikely that monotonicity holds. Specifically, some individuals may become irritated with the door-to-door canvassing, developing a more negative attitude towards transgender people in response, while their attitudes may have remained level or improved slightly over-time otherwise.}

\section{Conclusions}

Investigators frequently face questions that require reasoning about treatment effects within subgroups defined by post-treatment characteristics, such as implementation quality, compliance, and attentiveness. Yet, directly conditioning on post-treatment variables introduces problematic biases (e.g. \citealp{montgomery2018conditioning}). The TRACE provides a principled way to partially identify treatment effects for the subgroup that, if treated, would have taken on a particular value of a relevant post-treatment variable. In comparison to existing approaches to utilizing post-treatment variables, ours does not require strong and untestable assumptions such as the absence of mediator-outcome confounders or the exclusion restriction. Indeed, the TRACE is especially useful when the exclusion restriction not only fails, but where the direct effect that would violate it is central to the research question. Additionally, the TRACE is possible to use in cases where the relevant post-treatment variable is or can only be measured in the treatment group. 

The primary limitation of this approach is that it offers only partial identification, excepting the special case where we can argue for no effect in the ``non-reactive'' group ($\text{TRACE}(0)=0$). We emphasize that it is neither necessary nor desirable for users to postulate and defend a single value of TRACE(0). Instead, the results show what TRACE values are possible given beliefs we can defend about TRACE(0), together with restrictions revealed by the trimming bounds. For example, in some cases investigators may be able to argue that the TRACE(0) is near zero, exactly zero, has a smaller absolute value but same sign as, or has the opposite sign as that of the TRACE. A defensible assumption of this kind may sometimes lead to an informative boundary; in other cases it may not.

One area of particular interest may be the interpretation of null results from experimental studies in which implementation was uneven. For example, finding no average effects of community policing on citizen-police trust, cooperation with police, or crime across multiple countries, \citet{blair_community_2021} 
reasonably suggest that ``implementation challenges common to police reforms may have contributed to these disappointing results.” In such cases, the approach we propose enables researchers to precisely characterize what we would have to assume in order to attribute null effects to implementation challenges. 

%TC:ignore

%\paragraph{Replication Note from the Editors.}

\paragraph{Data Availability Statement.}
Replication code for this article has been published in the Political Analysis Harvard Dataverse at \url{https://doi.org/10.7910/DVN/HLLLGF} \citep{repdata}.

\paragraph{Acknowledgments.}
For helpful feedback on earlier versions of this paper, we thank Katherine Casey, Dean Knox, Shiyao Liu, Benjamin Morse, Molly Offer-Westort, and Anna Wilke, workshop participants at UCLA, UC Santa Barbara, and Vanderbilt University, and conference participants at the 2025 ACIC annual meeting in Detroit, EuroCIM 2025 in Ghent, the 2025 Polmeth annual summer meeting in Atlanta, and the 2025 APSA annual meeting in Vancouver. Mathew Boswell provided excellent research assistance. All errors are our own. The authors used Claude (Anthropic) and ChatGPT (OpenAI; standard conversational interface) for copyediting (e.g., identifying typographical errors and suggesting wording improvements to reduce length), to locate and compute numerical values from the \cite{knox_administrative_2020} replication archive used in~\ref{app:klmcomp}, and to assist in the preparation of the replication archive (including debugging, verifying code execution across computing environments, and checking adherence to journal guidelines). The authors reviewed all outputs and are responsible for all content.

\paragraph{Funding Statement.}
The authors declare that no specific funding was received for this article.

\paragraph{Conflicts of Interest.}
The authors declare none.

\paragraph{Ethical Standards.}
This study does not involve original human subjects research or primary data collection. All data used are from publicly available sources, and all analyses were conducted in accordance with applicable ethical standards.

%\paragraph{Declaration of AI Use.}
%The authors used Claude (Anthropic) and ChatGPT (OpenAI; standard conversational interface) for copyediting (e.g., identifying typographical errors and suggesting wording improvements to reduce length), to locate and compute numerical values from the \cite{knox_administrative_2020} replication archive used in~\ref{app:klmcomp}, and to assist in the preparation of the replication archive (including debugging, verifying code execution across computing environments, and checking adherence to journal guidelines). The authors reviewed all outputs and are responsible for all content.

%TC:ignore
\printbibliography
\clearpage
%TC:ignore
\appendix

\section{Equivalent alternative  quantities to reason about}
\label{alt-quant}

While we regard the TRACE(0) as entirely unknown, it has two components, one of which ($\E[Y(1) \mid M(1) = 0]$ ) can be identified (by $E[Y(1) \mid D = 1, M = 0]$) under randomization. The difficulty comes with the second component, $\E[Y(0)\mid M(1) = 0]$. Randomization allows this to be written as  $\E[Y(0) \mid D = 1, M = 0]$, but it remains unidentified. Assuming no defiers, $\E[Y(0) \mid D = 1, M = 0]$ is simply the expected non-treatment outcome among never-takers ("NT"), $\E[Y(0) \mid \text{NT}]$. Thus, one has the choice of either reasoning directly about $\E[Y(0) \mid \text{NT}]$, or reasoning about the entirety of TRACE(0). 

Second, one could also choose to reason about the average non-treatment outcome among compliers. This is because we observe an estimate of $\E[Y(0) \mid D = 0, M = 0]$, which is composed of averages among compliers ("C") and never-takers ("NT"), 
\begin{align}
    \E[Y(0) \mid D = 0, M = 0] &= \E[Y(0) \mid \text{ NT}] \frac{Pr(\text{NT})}{Pr(\text{NT}) 
    + Pr(\text{C})} \\
    &+ \E[Y(0) \mid \text{ C}] \frac{Pr(\text{C})}{Pr(\text{NT}) + Pr(\text{C})} \nonumber
\end{align}

\noindent Having observed the left-hand side, assuming a value for either $\E[Y(0) \mid \text{C}]$ or $\E[Y(0) \mid \text{NT}]$ is enough to fix the other. That is, one could make an assumption on $\E[Y(0) \mid \text{ C}]$, use the value of $\E[Y(0) \mid D = 0, M = 0]$ to back-out $\E[Y(0) \mid \text{NT}]$, and use that in turn to compute the TRACE(0) and identify the TRACE. 

This leaves us with a choice among three different quantities that we could reason about in order to get a range of estimates for TRACE: TRACE(0), $\E[Y(0) \mid \text{NT}]$, or $\E[Y(0) \mid \text{C}]$. Note, however, that using the latter two quantities depends on an assumption of no defiers, which is not required if we directly reason about TRACE(0). All three of these routes to (partial) identification are equivalent, but some may be better suited to reasoning and argumentation than others in a given context.  In the cases examined here, we found it profitable to reason directly about TRACE(0) because it refers to a substantive causal effect in a sub-group, and the nature of this sub-group can support arguments about TRACE(0), e.g. that is might have the opposite sign as the TRACE, that it is likely to be zero, negative, smaller than the effect in the other group, or something else useful for bounding.

\section{Latent DAG representation}
\label{app.dag}

\begin{figure}[htp!]
\begin{center}
\begin{tikzpicture}[->, thick]
\node (d) at (0,0) {$D$};
\node (y) at (4,0) {$Y$};
\node (m) at (2,0) {$M$};
\node (n) at (2,1) {$M(1), M(0)$};
\node (u) at (4.5,2) {$U$};
\node (x) at (2.5,2) {$X$};
\path (m) edge (y);
\path (n) edge (m);
\path (d) edge (m);
\draw (d) to [bend right=30] (y);
\path[dashed] (u) edge (n);
\path[dashed] (u) edge (y);
%\path[dashed] (u) edge (m);
\path (x) edge (n);
\path (x) edge (y);
\end{tikzpicture}
\caption{Latent DAG representation}\label{fig:mstar-other-dag}
\end{center}
\end{figure}
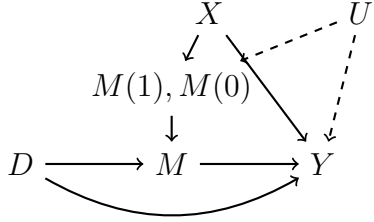

One difficulty of using DAGs in this context is that the estimand we propose conditions on membership in a group---those units with $M(1)=1$---that cannot be conditioned on using observed variables, and thus cannot be directly represented by a conditioning procedure on the DAG. To represent our estimand thus requires modifying what appears on the DAG.  Specifically, consider the value of $\{M(1), M(0)\}$ for each unit $i$ as a (latent) variable. The characteristic $M(1) = 1$ represents a latent ``type'', or membership in either of two principal strata---that with $\{ M(1)=1, M(0)=1\}$ (or ``always-takers'') and that with $\{M(1)=1, M(0)=0\}$ (``compliers''). The value of $M(1)$ is a function of the unobservables, $U$, and can therefore be represented as a descendant of $U$ on the graph. The realized $M=M(d)$ is a (deterministic) function of $M(1)$, $M(0)$, and $D$. These relationships are encoded in Figure~\ref{fig:mstar-other-dag}. A key feature of this relationship is that there is no $D \to M(1)$ edge, so conditioning on $M(1)$ (and/or $M(0)$) does not jeopardize identification of the $D \to Y$ effect.

\section{Applicability to partial identification using AutoBounds}\label{app:autobounds}

\citet{duarte2024automated} describe a general framework for obtaining bounds on causal estimands by solving a polynomial optimization problem. When formulated as minimization and maximization problems, these programs yield lower and upper bounds on causal queries subject to assumptions and observed data. Although the TRACE is not among the estimands considered explicitly in \citet{duarte2024automated}, users could treat the TRACE as the target estimand and encode postulated values or constraints on TRACE(0). AutoBounds would then provide a convenient way to obtain bounds on the TRACE, and to combine the assumptions used here with any additional restrictions an investigator is willing to impose, revealing their joint identifying power. 

Suppose $Y$ is binary. Then, the TRACE can be written as follows:
\begin{align}
    TRACE = \frac{Pr(Y(D=1) = 1, M(D=1) = 1)}{Pr(M(D=1)=1)} - \frac{Pr(Y(D=0) = 1, M(D=1) = 1)}{Pr(M(D=1)=1)}
\end{align}

This can then be written as a polynomial objective function following the approach for the polynomialization of the LATE described in \cite{cooper-autobounds} (Appendix E.1). We can also polynomialize modeling assumptions on TRACE(0). Writing TRACE(0) in the same form as we did the TRACE, we can formulate a constraint
\begin{align}
    \frac{Pr(Y(D=1) = 1, M(D=1) = 0)}{Pr(M(D=1)=0)} - \frac{Pr(Y(D=0) = 1, M(D=1) = 0)}{Pr(M(D=1)=0)} < C
\end{align}
where $C$ is an assumed bound on value of TRACE(0). This expression can be simplified to its polynomial form according to \cite{duarte2024automated}, and used to write constraints based on lower and upper bounds on the value of TRACE(0). We leave the implementation of this approach to future work. However as we have shown that the no-assumption trimming bounds are sharp, we conjecture that solving the optimization problem using the outlined objective function and constraints, in combination with the constraints imposed by the AutoBounds algorithm, will yield the same bounds that we arrive at analytically here.

\section{Equivalence of trimming bounds on TRACE(0) and TRACE}
\label{app:samebounds}

Here we show algebraically how application of trimming bounds on TRACE or on TRACE(0) are identical. To work entirely with sample quantities, define $\overline{Y(1)}$ and
$\overline{Y(0)}$ as the sample means of $Y$ among treated and control units,
respectively (i.e., observed outcomes, not potential outcomes over all units). 

Let $\overline{Y(0)}_{\text{low}M11}$ and
$\overline{Y(0)}_{\text{top}M11}$ denote the average outcomes among controls
in the lowest and highest fractions, respectively, of size
$M_{11}=\widehat{\Pr}(M(1)=1)$ of the control outcome distribution.
Similarly, define
$\overline{Y(0)}_{\text{low}M10}$ and
$\overline{Y(0)}_{\text{top}M10}$ as the averages among controls in the
lowest and highest fractions of size
$M_{10}=\widehat{\Pr}(M(1)=0)=1-\widehat{\Pr}(M(1)=1)$.

Recalling that the TRACE estimator is
\begin{align*}
\widehat{TRACE}
&=
\frac{\overline{Y(1)}-\overline{Y(0)}
-\widehat{TRACE(0)}(1-M_{11})}{M_{11}},
\end{align*}
we now determine what value of TRACE corresponds to imposing the lower
trimming bound on TRACE(0).

\begin{align}
M_{11}\,\widehat{TRACE}
&= \overline{Y(1)}-\overline{Y(0)}
   -\widehat{TRACE(0)}_{\text{low}}(1-M_{11}) \\
&=
\underbrace{\overline{Y(1)}
-\widehat{\E}[Y(1)\mid M(1)=0](1-M_{11})}
_{=\,M_{11}\widehat{\E}[Y(1)\mid M(1)=1]}
-\overline{Y(0)}
+\overline{Y(0)}_{\text{top}M10}(1-M_{11}) \\
&=
M_{11}\widehat{\E}[Y(1)\mid M(1)=1]
-\overline{Y(0)}
+\overline{Y(0)}_{\text{top}M10}(1-M_{11}) \\
&=
M_{11}\widehat{\E}[Y(1)\mid M(1)=1]
-\underbrace{\bigl(
\overline{Y(0)}
-\overline{Y(0)}_{\text{top}M10}(1-M_{11})
\bigr)}
_{=\overline{Y(0)}_{\text{low}M11}M_{11}} \\
&=
M_{11}\widehat{\E}[Y(1)\mid M(1)=1]
-\overline{Y(0)}_{\text{low}M11}M_{11}.
\end{align}

Dividing by $M_{11}$ gives
\begin{equation}
\widehat{TRACE} =
\widehat{\E}[Y(1)\mid M(1)=1]
-\overline{Y(0)}_{\text{low}M11}
=
TRACE_{\text{high}}.
\end{equation}

Likewise, imposing the upper trimming bound on $TRACE(0)$ produces
$TRACE_{\text{low}}$. Hence, the no-assumption trimming bounds on
$TRACE(0)$ imply exactly the no-assumption trimming bounds on TRACE in
every sample.

\section{Relationship of the TRACE to the difference in conditional means}\label{app:naivecomp}

Here, we show the relationship between the TRACE and the difference in conditional means, or the ``naive'' comparison that would be made by directly conditioning on an observed value of $M$, $\text{DIM}_{\text{M=1}} \equiv \E[Y \mid D=1, M = 1]-\E[Y \mid D=0, M = 1]$.

This is the quantity often used when investigators drop observations based on a post-treatment criterion, such as passing an attention screen or a manipulation check \citep{montgomery2018conditioning}. It is also relevant in cases where $M$ represents selection into the data, as in the police violence application discussed in Section~\ref{subsec.app1} above (see also \cite{knox_administrative_2020}).

Consider the following breakdown of this estimator, where ``AT'', ``C'', and ``Def'' refer to always-takers, compliers, and defiers respectively,
\begin{align}
\text{DIM}_{\text{M=1}} &\equiv \E[Y_i \mid \;D_i=1,M_i=1]-\E[Y \mid D_i=0,M_i=1] \\
&=  \E[Y_i(1) \mid M_i(1)=1] - \E[Y_i(0) \mid M_i(0) = 1]  \\
&=  \E[Y_i(1) \mid M_i(1)=1] - \E[Y_i(0) \mid \text{Def+AT}]  \\
&= \E[Y_i(1) \mid M_i(1)=1] - \E[Y_i(0) \mid M_i(1)=1] + \E[Y_i(0) \mid M_i(1)=1] - \E[Y_i(0) \mid \text{Def+AT}] \\
&= TRACE + \E[Y_i(0) \mid \text{AT+C}] - \E[Y_i(0) \mid \text{Def+AT}] \label{bias}
\end{align}

\noindent Under monotonicity, we are left with 
\begin{align}
\text{DIM}_{\text{M=1}} &=  TRACE + \E[Y_i(0) \mid \text{AT+C}] - \E[Y_i(0) \mid \text{AT}] \\
&=  TRACE + \left[ \E[Y_i(0) \mid \text{AT}]\frac{Pr(\text{AT})}{Pr(\text{AT})+Pr(\text{C})} + \E[Y_i(0) \mid \text{C}]\frac{Pr(\text{C})}{Pr(\text{AT})+Pr(\text{C})}\right] - \E[Y_i(0) \mid \text{AT}] \nonumber \\
&=  TRACE + \left[ \E[Y_i(0) \mid \text{C}]-\E[Y_i(0) \mid \text{AT}] \right] \frac{Pr(\text{C})}{Pr(\text{AT})+Pr(\text{C})}\label{bias.2}
\end{align}

\subsection{Bounding TRACE from the naive comparison: application to the police stops example}\label{app:klmcomp}

As noted in Section~\ref{subsec.app1} in the main text, in the police violence setting -- where $M$ is a ``necessary condition mediator'' and when we are additionally willing to assume monotonicity of $D$ in $M$ -- Equation~\ref{bias.2} leads to a bounding option for TRACE. Practically, this means we can bound the TRACE only using data from observed police stops.

Here, in the absence of data on encounters that do not result in stops, $\frac{Pr(C)}{Pr(C)+Pr(AT)}$ will be unknown. However, since it is between $0$ and $1$, 
\begin{equation}
    \text{DIM}_{M=1} \leq \text{TRACE} \leq \text{DIM}_{M=1}+\E[Y \mid D=0,M=1].
\end{equation}

\noindent Further, recalling that $\text{DIM}_{\text{M=1}} = \E[Y \mid D=1,M=1]-\E[Y \mid D=0,M=1]$, this is simply
\begin{equation}
    \E[Y \mid D=1,M=1]-\E[Y \mid D=0,M=1] \leq \text{TRACE} \leq \E[Y \mid D=1,M=1].
\end{equation}

\noindent This form also suggests an easy intuition: $\E[Y \mid D=1,M=1]$ is the mean $Y(1)$ among always-takers and compliers. To obtain the TRACE, we need to subtract $\E[Y(0) \mid \text{AT}+\text{C}]$, which is not identified when we only observe units with $M=1$. However, this quantity is at least zero, producing an upper bound on the TRACE of $\E[Y \mid D=1,M=1]$. At the other extreme, we know that $\E[Y(0) \mid D=0, M=1] \geq \E[Y(0) \mid \text{AT}+\text{C}]$, because $Y(0)$ must be equal to zero for compliers.

As applied to  \cite{knox_administrative_2020}, we can obtain these quantities directly from replication data (\url{https://doi.org/10.7910/DVN/KFQOCV}), the 
\verb|sqf_fryer_2003_2013.Rdata|
file).

\begin{lstlisting}
load("sqf_fryer_2003_2013.Rdata")
d <- d2; rm(d2)
d$race <- factor(d$race2,
                 levels = c("white","black","hisp","asian","other"),
                 labels = c("white","black","hispanic","asian","other"))
# E[Y | D=0, M=1]: use-of-force rate among stopped Whites
ey_white <- mean(d$force2[d$race == "white"], na.rm = TRUE)
# = 0.1726

# E[Y | D=1, M=1]: force rate among stopped minorities
ey_black <- mean(d$force2[d$race == "black"], na.rm = TRUE)
# = 0.2334
ey_hisp <- mean(d$force2[d$race == "hispanic"], na.rm = TRUE)
# = 0.2394

# DIM_{M=1}
# Black vs White: 0.2334 - 0.1726 = 0.0608
# Hispanic vs White: 0.2394 - 0.1726 = 0.0668
\end{lstlisting}

Accordingly, the bounds obtained when comparing drivers whose police-perceived race was Black vs. White are 6.1\% to 23.3\%. Comparing drivers perceived to be Hispanic vs. White, the bounds are 6.7\% to 23.9\%. These cover the point estimates in \citet{knox_administrative_2020}, which they produce using approximate estimates of $\frac{Pr(\text{C})}{Pr(\text{C})+Pr(\text{AT})}$ based on auxiliary data.

\begin{comment}

\end{comment}
%TC:endignore

\end{document}